\documentclass[a4paper,nofootinbib,10pt]{revtex4}

\usepackage[utf8]{inputenc}
\usepackage{amsmath}
\usepackage{graphicx}
\usepackage{caption}
\usepackage{amssymb}
\usepackage{color}
\usepackage{cancel}
\usepackage{framed}
\usepackage{comment}
\usepackage{mathtools}
\usepackage{hyperref}
\usepackage{chngcntr}
\usepackage{wasysym}

\counterwithin{equation}{section}

\def\bea{\begin{eqnarray}}
\def\eea{\end{eqnarray}}
 
 \newcommand{\beginsupplement}{%
 	\setcounter{section}{0}
        \renewcommand{\thesection}{S\arabic{section}}%
        \setcounter{table}{0}
        \renewcommand{\thetable}{S\arabic{table}}%
        \setcounter{figure}{0}
        \renewcommand{\thefigure}{S\arabic{figure}}%
     }

\def\Mp{M_{\rm Pl}}
\def\Mpl{M_{\rm Pl}}

\def\TeV{\,{\rm TeV}}
\def\GeV{\,{\rm GeV}}

\def\eV{\,{\rm eV}}

\newcommand{\vev}[1]{\langle #1 \rangle}

\newcommand{\nn}{\nonumber}
\newcommand{\pd}{\partial}

\def\a{\alpha}  \def\g{\gamma} 
\def\d{\delta}
\def\g{\gamma} \def\m{\mu} \def\n{\nu}  
   
 \def\G{\Gamma}  \def\L{\Lambda}

 \newcommand{\Ocal}{{\mathcal O}}
 
 \newcommand{\Lcal}{{\mathcal L}}

\newcommand*{\affaddr}[1]{#1} 
\newcommand*{\affmark}[1][*]{\textsuperscript{#1}}

\def\Weizmann{\small{Department of Particle Physics and Astrophysics, Weizmann Institute of Science, Rehovot 761001, Israel}}

\def\Mainz{\small{Helmholtz Institute Mainz, Johannes Gutenberg University, Mainz 55099, Germany}}
\def\Berkeley{\small{Department of Physics, University of California, California 94720-7300, USA}}

\begin{document}

\title{\Large\bfseries Relaxion Stars and their detection via Atomic Physics}

\author{
Abhishek Banerjee\affmark[$\oplus$], 
Dmitry Budker\affmark[$\odot$]\affmark[$\!\!\rightmoon$], 
Joshua Eby\affmark[$\oplus$], 
Hyungjin Kim\affmark[$\oplus$], 
and Gilad Perez\affmark[$\oplus$]
\vspace*{.2cm}\\
{\it\affaddr{\affmark[$\oplus$]\Weizmann}}\\
{\it\affaddr{\affmark[$\odot$]\Mainz}}\\
{\it\affaddr{\affmark[$\rightmoon$]\Berkeley}}\\
}

\begin{abstract}
The cosmological relaxion can address the hierarchy problem, while its coherent oscillations can constitute dark matter in the present universe. We consider the possibility that the relaxion forms gravitationally bound objects that we denote as relaxion stars. The density of these stars would be higher than that of the local dark matter density, resulting in enhanced signals in table-top detectors, among others. Furthermore, we raise the possibility that these objects may be trapped by an external gravitational potential, such as that of the Earth or the Sun. This leads to formation of \emph{relaxion halos} of even greater density. We discuss several interesting implications of relaxion halos, as well as detection strategies to probe them.
\end{abstract}

\maketitle

\section{Introduction} 
Resolving the nature of the dark matter (DM) is one of the most fundamental questions in modern physics~\cite{Bertone:2016nfn}.
Although particle DM at the electroweak scale is a highly motivated solution~\cite{Jungman:1995df}, no discovery of such DM was made to date, either directly~\cite{Bertone:2004pz,daSilva:2017swg,Aprile:2018dbl}, indirectly~\cite{Gaskins:2016cha} or at the LHC~\cite{Boveia:2018yeb}. Another intriguing possibility is that of a cold, ultra-light, DM field, coherently oscillating to account for the observed DM density. We consider a class of models where a light scalar particle composes the DM. A well-motivated example is the relaxion, where even a minimal model that addresses the hierarchy problem~\cite{Graham:2015cka} may lead to the right relic abundance in a manner similar to axion models, however geared with a dynamical misalignment mechanism~\cite{Banerjee:2018xmn} for relaxion masses roughly above $10^{-11}\eV$.
Due to spontaneous CP violation, the relaxion mixes with the Higgs, and, as a result, acquires both pseudoscalar and scalar couplings to the Standard Model (SM) fields~\cite{Flacke:2016szy,Choi:2016luu} (this effect could be suppressed in particle-production-based models~\cite{Hook:2016mqo}). The latter distinguishes the relaxion from axion dark matter, which has only pseudoscalar couplings, and where the same property of generation of CP violation was shown to lead to a solution of the strong CP problem~\cite{Davidi:2017gir} as well as potentially generating the cosmological baryon asymmetry~\cite{Abel:2018fqg}.

A striking consequence of the relaxion-Higgs mixing is that, as the relaxion forms a classical oscillating DM background, all basic constants of nature effectively vary with time since they all depend on the Higgs vacuum expectation value~\cite{Banerjee:2018xmn}.
(For earlier discussion in the context of dilaton DM see~\cite{Arvanitaki:2014faa,Graham:2015ifn,Safronova:2017xyt}.)
There are active experimental efforts searching for this form of scalar DM ({\it e.g.}~\cite{VanTilburg:2015oza,Hees:2016gop,Geraci:2018fax, Aharony:2019iad,Stadnik:2016zkf,Stadnik:2015kia,Rosenband1808}).
Despite the unprecedented accuracy achieved by the various searches, none of the current experiments reach the sensitivity required to probe physically motivated models. Furthermore, the resulting sensitivity in the region of our main interest, characterised by oscillation frequencies above the Hz level, is weaker than that of the probes related to fifth-force searches and equivalence-principle tests~(see {\it e.g.}~\cite{Arvanitaki:2014faa,Graham:2015ifn,Geraci:2018fax,Safronova:2017xyt,Arvanitaki:2015iga,Aharony:2019iad,Frugiuele:2018coc,Flacke:2016szy,Hees:2018fpg,Rosenband1808}).

In this paper, we demonstrate that if the scalar DM forms a self-gravitating compact object, usually known as a \emph{boson star}, its density would be higher than that of the local DM density, resulting in enhanced signals for table-top detectors, among others.
Furthermore, we raise the possibility that these objects may be trapped by the gravitational potential of the Earth or the Sun.
This leads to formation of a \emph{relaxion halo} with a much larger density, compared to that of local DM.
We discuss several interesting implications and also detection strategies that are presented below. We work in natural units, where $\hbar = c = 1$.

\section{Coherent dark matter background} \label{sec:DM}

For concreteness, among all possible relaxion couplings to SM particles, we focus on the following interactions:
\begin{equation} \label{LagYuk}
 \Lcal \supset g_{e} \phi\,\bar{e}\,e + \frac{g_{ \g}}{4} \phi\,F_{\m\n}F^{\m\n},
\end{equation}
where $\phi$ is the scalar DM field, $e$ is the electron field, and $F^{\m\n}$ is the electromagnetic field strength. 
The oscillation of the scalar field induces an oscillation in the electron mass, $m_e$, and the fine structure constant, $\alpha$, with frequency $\omega \approx m_\phi$.
For this reason, atomic precision measurements looking for variation of fundamental constants can probe models of scalar-field dark matter.
For the following discussion, we take a phenomenological approach and consider $g_e$ and $g_\gamma$ as independent parameters (see~\cite{Gupta:2015uea,Davidi:2018sii} for possible microscopic origins of these couplings). 

A concrete instantiation of this scenario is the relaxion, which has a potential of the form \cite{Graham:2015cka,Banerjee:2018xmn,Flacke:2016szy}
\begin{equation}
 V(H,\phi) = \left(\Lambda^2 - g\,\Lambda\,\phi\right) \left|{H}\right|^2 - c\,g\,\L^3\,\phi - \frac{\L_{\rm br}^4}{v^2}\left|H\right|^2\,\cos\frac{\phi}{f},
\end{equation}
where $\L$ is the cutoff scale for the Higgs mass, $g \sim \L_{\rm br}^4/f\,\L^3$ is a dimensionless coupling parameter, $c$ is an $\Ocal(1)$ coefficient, $\L_{\rm br}$ is the backreaction scale, and $v$ is the electroweak scale. In this proposal, the rolling of the field $\phi$ due to the linear term dynamically scans the Higgs mass parameter, until eventually the backreaction potential stops the rolling when $\vev{H} = \Ocal(v)$, solving the electroweak hierarchy problem~\cite{Graham:2015cka}. It was shown recently that with minimal additional assumption about the inflation sector, such a relaxion naturally makes a viable DM candidate~\cite{Banerjee:2018xmn}. The DM energy density is generated by the misalignment mechanism after the rolling stops, from coherent oscillations of the field around its minimum generated during reheating. The model dependence can be simplified by parameterizing the theory in terms of $T_{\rm ra}$, the temperature at which the backreaction potential reappears after reheating.

For the relaxion model (or other Higgs portal-like theories), scalar couplings to matter are generated by mixing with the Higgs~\cite{Flacke:2016szy,Choi:2016luu}, and so can be parameterized by a mixing angle $\sin\theta$; for the couplings of Eq.~\eqref{var_const}, one has $g_e = y_e\,\sin\theta$ and $g_\g \sim (\a / 4\pi v)\,\sin\theta$, where $y_e$ is the Higgs Yukawa coupling to the electron. By generic naturalness arguments, one may additionally require $g_e \lesssim 4\pi m_\phi/\Lambda$. We will use this model as a benchmark for comparison, though our conclusions will hold more generally for many forms of light scalar dark matter.

To investigate whether the variation of fundamental constants induced by the $\phi$-oscillation is measurable, we must compute variations of fundamental constants in terms of the model parameters,
\bea 
\frac{\delta m_e}{\langle m_e\rangle} = \frac{g_e \phi}{\langle m_e\rangle}\,,
\qquad
\frac{\delta \alpha}{\alpha} = g_\gamma \phi\,,
\label{var_const}
\eea
where $\langle m_e\rangle$ corresponds to the time-averaged electron mass (see discussion in~\cite{Kozlov:2018qid,Aharony:2019iad}). 
Given the experimental sensitivity to $\delta m_e /\langle m_e\rangle$ and $\delta \alpha /\alpha$, and also the amplitude of the $\phi$-oscillation in a given model, we can estimate the sensitivity to $g_e$ and $g_\gamma$.

For a light scalar field with $m_\phi \gtrsim 10^{-10}$ eV, there has been in a blind spot for experimental measurements of time variations of fundamental constants (see~\cite{Safronova:2017xyt} for a recent review). In~\cite{Aharony:2019iad}, using dynamical decoupling with trapped ions  resulted in a bound on scalar particle masses in the range $m_\phi\sim 10^{-11}-10^{-10}\,$eV (roughly $1-10$ kHz oscillation frequency) with accuracy of $1:10^{13-14}$ for both
$\delta m_e / \langle m_e\rangle$  and $ \delta \alpha/\alpha\,.$ 
The bound was obtained via atom-cavity comparison~\cite{Wcislo:2016}, where for $\delta m_e / \langle m_e\rangle$, this method can only be effectively used for frequencies $\gtrsim 10\,$kHz~\cite{Geraci:2018fax}. These bounds can be improved by roughly two orders of magnitude and can cover the range up to 10 MHz. A broader range of masses corresponding to frequencies up to 100 MHz can be covered using conventional Doppler-free techniques such as  polarization spectroscopy, using optical transitions in atoms and molecules contained in vapor cells. Assuming one year total of interrogation time can effectively bring the sensitivity to roughly $1:10^{18}\,$~\cite{Antypas:2019qji}.

At smaller masses $m_\phi \lesssim 10^{-13}$ eV, the best bounds on $\delta m_e / \langle m_e\rangle$  arise from atomic-clock comparisons between hyperfine and optical transitions, which have a relative projected accuracy of roughly $1:10^{16}$ where the hyperfine clock uncertainty is saturated (see, for example,~\cite{6174184}). 
As for $\delta \alpha / \alpha$, different atomic-clock comparisons~\cite{RevModPhys.87.637} as well as measurements of special ``forbidden" transitions in highly charged ions to optical transitions can reach accuracies of roughly $1:10^{18-19}$~\cite{PhysRevA.92.060502, JunYe2019}.

If this scalar coherent oscillation corresponds to dark matter in our local neighbourhood, the amplitude is fixed. 
It is given, within a coherent patch, as (see {\it e.g.}~\cite{Arvanitaki:2014faa,Arvanitaki:2015iga})
\bea
\phi(t) = \frac{\sqrt{2\rho_{\rm local}}}{m_\phi} \sin(m_\phi t)
	= 3\times 10^{-3} \eV  \times \left( \frac{1 \eV}{m_\phi} \right)  \sin(m_\phi t)\,,
\label{phi_dm}
\eea
where we take $\rho_{\rm local} = 0.4\GeV/{\rm cm}^3$ as the local dark matter density. 
Various theoretical and experimental efforts have been put forward to probe effective variation of fundamental constants induced by a coherently oscillating background DM field. 
As it can be seen from Eq.~\eqref{phi_dm}, 
the effect is strongest when the mass is the lightest, $m_\phi\simeq 10^{-21}\eV$, which is marginally allowed by the observation of large-scale structures of the universe~\cite{Irsic:2017yje,Armengaud:2017nkf} or measured rotational velocities in galaxies \cite{Bar:2018acw}. 
Substituting this expression to Eq.~\eqref{var_const}, one can compute the variation of fundamental constants, but the resulting effect is small; in the range $m_\phi \gtrsim 10^{-15}$ eV, the sensitivity estimates discussed above suggest it is difficult to compete with the bounds that arise from fifth-force experiments~\cite{VanTilburg:2015oza,Hees:2016gop,Rosenband1808}.
At smaller masses $10^{-21}\eV \leq m_\phi \leq 10^{-15}\eV$, atomic-clock comparison tests (see {\it e.g.}~\cite{Safronova:2017xyt} and Refs. therein) can compete with or be stronger than fifth-force constraints, though this range does not overlap with the region of relaxion DM models~\cite{Banerjee:2018xmn}.

\section{Relaxion Stars} \label{sec:stars}
In this section, we consider the case where the scalar DM forms a bound state with much larger density compared to background dark matter, due to its own self-gravity and self-interactions.
These are typically known as boson stars or axion stars (or in the more specific case, \emph{relaxion stars}).
Here, we investigate whether atomic precision measurements can probe the existence of such compact objects when they pass through the Earth.
A boson star is described by a classical scalar field, oscillating coherently with frequency approximately equal to the scalar particle mass.
Similar to the discussion above, a crucial quantity for precision measurement is the amplitude of oscillation, $\phi = \sqrt{2\rho_\star}/ m_\phi $, which is determined by the density $\rho_\star$ of the compact object. Note that we have dropped the explicit time dependence of $\phi$ for notational simplicity, and will from now on take $\phi$ as the amplitude and $m_\phi$ as the frequency of oscillation.

Formation of boson stars is a rapidly evolving field. In the context of QCD axions, overdensities known as miniclusters can be produced on small physical scales (compared to the scale of galaxies) if the Peccei-Quinn symmetry is broken after inflation \cite{Kolb:1993zz}. These miniclusters have been recently shown to form self-gravitating boson stars on short timescales (compared to galaxy lifetimes), through a process described by gravitational relaxation of effective quasiparticles \cite{Hui:2016ltb,Bar-Or:2018pxz}; this has been investigated assuming both idealized initial conditions \cite{Levkov:2018kau}, as well as more realistic ones \cite{Eggemeier:2019jsu} determined by large-scale simulations of QCD axions \cite{Vaquero:2018tib}. For ultralight axions, large-scale simulations also show boson star-like objects forming in the central cores of galaxies \cite{Schive:2014dra,Veltmaat:2018dfz}, suggesting that this is a generic property of light scalar field dark matter. The spectrum of initial density fluctuations has not similarly been investigated for relaxions, a topic we delay for future work; for the present purposes, we merely point out that a similar formation mechanism might hold for boson stars formed from relaxions as well. 

\begin{figure}[t]
\centering
\includegraphics[scale=.305]{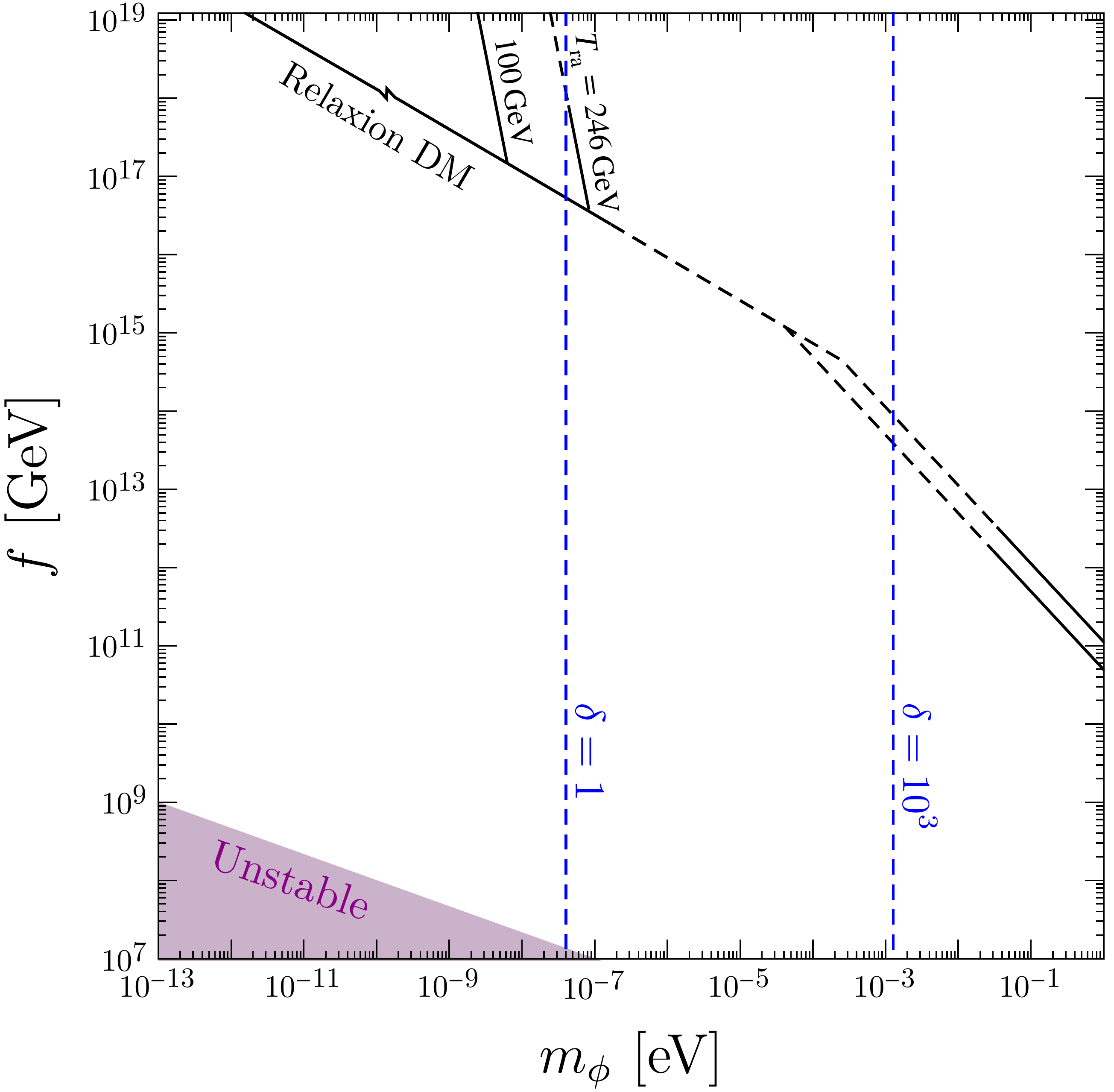}
\qquad
\includegraphics[scale=.29]{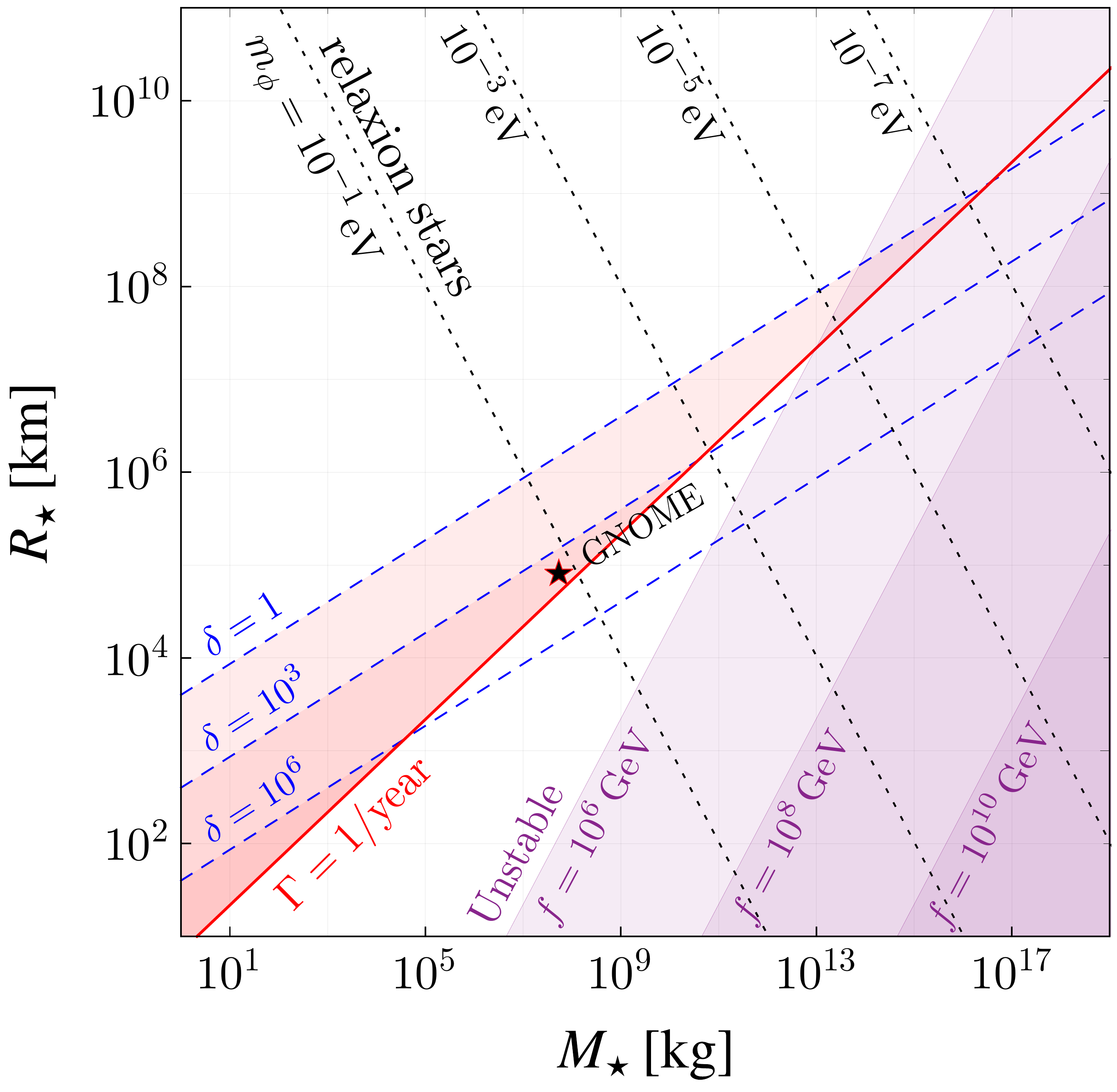}
\caption{%
The relevant parameter space for transient DM boson stars encountering the Earth. In both panels, the dashed blue lines are contours of constant overdensity $\d$, and the purple shaded regions indicate instability through self-interactions.
Left: parameter space in scalar mass $m_\phi$ and decay constant $f$ allowing for gravitationally stable objects, assuming the encounter rate is $\G=1/$year. The black lines denote coherent relaxion DM for different choices of reappearance temperature $T_{\rm ra}$, as discussed in Section \ref{sec:DM}; solid lines are allowed parameters, and dashed are ruled out by fifth-force constraints (see \cite{Banerjee:2018xmn} for further information).
Right: $M_\star$ and $R_\star$ are treated as independent parameters; the black dotted lines denote stable configurations formed from scalars of mass $m_\phi$, and
the red shaded region represents $\G>1/$year and $\d > 1$. The black star represents the benchmark point used by the GNOME collaboration~\cite{JacksonKimball:2017qgk}.
}
\label{fig:transient}
\end{figure}

A compact object is independent of background dark matter and its density does not necessarily coincide with that of the background.
In the presence of gravity, a free scalar field can support itself against collapse through repulsive gradient energy (that is, effective pressure sourced by the kinetic energy of the field); this leads to a unique relation between its radius $R_\star$ and mass $M_\star$,
\begin{equation} \label{Rstar}
 R_\star = \frac{\Mp^2}{m_\phi^2}\frac{2}{M_\star}\,,
\end{equation}
where $\Mp = 1.2 \times 10^{19}\GeV$ is the Planck mass. 
Some generic properties of boson stars are reviewed in the Supplementary Material \ref{app:ASts}.
The overdensity inside a boson star compared to the background density of DM would correspond to
\begin{align}
 \d \equiv \frac{\rho_\star}{\rho_{\rm local}} = \frac{2\,\Mpl^2}{7\pi\,m_\phi^2\,R_\star^4}\frac{1}{\rho_{\rm local}} 
 		\approx 7\times10^{21}\left(\frac{10^{-10}\text{ eV}}{m_\phi}\right)^2\left(\frac{10^5\text{ km}}{R_\star}\right)^4,
\end{align}
where we used the approximate profile of Eq.~\eqref{app_profile} at $r\ll R_\star$ (see Supplementary Material), and Eq.~\eqref{Rstar}.
In this estimation, the benchmark choice for $m_\phi$ is consistent with the concrete relaxion DM model of~\cite{Banerjee:2018xmn}; in this case, we
would expect to gain a $\sqrt{\d} \approx 10^{11}$ enhancement in the amplitude of $\phi$ if such an object passes through the Earth. 
This leads to a relatively large effective variation of fundamental constants, compared to the case where such variation is induced by the standard background dark matter density.

However, the encounter rate between such stars and the Earth is low. 
To estimate how many such encounters would take place per year, we assume that an ${\cal O}(1)$ fraction of local dark matter is in the form of stable bound states with a fixed mass $M_\star$; the actual distribution of boson star masses depends critically on the formation history, which is beyond the scope of this paper. We also assume a geometric cross-section $\sigma_\star = \pi\,R_\star^2\,$, and that the motion of the boson stars obeys the virial relation in terms of their typical distribution, implying a speed of $v_\star = 10^{-3}$.
Under these assumptions, the encounter rate between the Earth and such objects is
\begin{align}
\G 
= n_\star \sigma_\star v_\star 
= \frac{\rho_{\rm local}}{M_\star}\pi\, R_\star^2\, v_\star
\approx 2\times10^{-18}\text{ yr}^{-1} \Big(\frac{m_\phi}{10^{-10}\text{ eV}}\Big)^2 \Big(\frac{R_\star}{10^5\text{ km}}\Big)^3.
\end{align}
From this estimate, we see that these encounters are so rare that an encounter typically does not occur during the entire history of the universe. More generally, the encounter rate increases with smaller $\delta$ as 
\begin{equation}
 \G \approx 0.05\text{ yr}^{-1} \times {\d^{-3/4}} \sqrt{\frac{m_\phi}{10^{-10}\text{ eV}}}.
\end{equation}

In the left panel of Fig.~\ref{fig:transient}, we identify the parameter space of relaxion mass $m_\phi$ and decay constant $f$ in which a collision rate $\G = 1$/year is possible. 
Although we ignore the self-interactions of relaxions, we include the decay constant $f$ in the plot to present a benchmark relaxion DM model (black solid lines allowed parameters, dashed ruled out by fifth-force meaurements)~\cite{Banerjee:2018xmn} and the region where the relaxion star is unstable due to the self-interaction (purple shaded region, see Eq.~\eqref{Mcritic} in Supplementary Material for details regarding the self-interaction potential).
The overdensity $\delta$ is also denoted by blue dashed lines, assuming the rate of one collision per year.
If $m_\phi\lesssim 10^{-8}$ eV, an overdensity $\d>1$ along with $\G = 1$/year is not possible for a self-gravitating object; 
only if $m_\phi > 10^{-8}\eV$ is this scenario viable.
This mass range corresponds to frequencies greater than order MHz, which can be probed using experimental techniques discussed in Section \ref{sec:DM}.%
\footnote{For the case of axionlike particle (ALP), it has been proposed that pseudo-scalar coupling to nucleons can be probed by using nuclear magnetic resonance techniques even when $m_\phi \gtrsim 10^{-8}\eV$ and $\delta =1$~\cite{JacksonKimball:2017elr}.
Although we do not discuss it in this paper, this experimental technique can equally apply to the scenario of transient relaxion stars since the relaxion could also have pseudo-scalar coupling to the SM fields.}

In the right panel of Fig.~\ref{fig:transient}, we show the mass-radius relation of boson stars (dotted lines).
Similar to the figure on the left panel, the purple shaded region denotes boson stars that are unstable to collapse due to self-interactions, while the blue dashed lines denote the density contrast $\delta = \rho_\star/\rho_{\rm local}$. 
The red shaded region represents $\d > 1$ and $\G > 1/$year, which is attainable only for $m_\phi \gtrsim 10^{-8}\eV$.
In other words, for scalar mass $m_\phi\lesssim 10^{-8}\eV$, it is either the case that the density of boson star is large but its encounter rate is too small for terresterial experiments, or that the rate is large enough but its density becomes even smaller than that of the background DM.
Note that possible transient signals induced by axion stars have already been investigated in~\cite{JacksonKimball:2017qgk}, where it is concluded that the Global Network of Optical Magnetometers for Exotic physics searches (GNOME) can probe ALP parameter space for $m_{\rm ALP} <10^{-13} \eV$, and that the projected sensitivity surpasses astrophysical constraints, which may seem to contradict Fig.~\ref{fig:transient}.
In~\cite{JacksonKimball:2017qgk}, the approach taken is more phenomenological, assuming $M_\star$ and $R_\star$ to be fully independent, which allows some region of parameter space to be probed by simultaneously satisfying $\G = \Ocal(1)$/year and $\d\gg 1$. 
This also indicates that the axion stars considered in~\cite{JacksonKimball:2017qgk} are not truly ground-state configurations.
We show the benchmark point used in~\cite{JacksonKimball:2017qgk} ($R_\star = 10R_\oplus$ and $M_\star = 4\times10^7$ kg) as the black star in the figure.

\section{Relaxion Halo} \label{sec:halo}
The formation of boson stars is a complex dynamical process. Typical investigations involve simulations of scalar-field dynamics, and commonly neglect any effect from baryons~\cite{Kolb:1993zz,Levkov:2018kau,Vaquero:2018tib}. 
In this section we suggest that, in the presence of baryons, gravitational relaxation may lead to configurations in which a large density of scalar field becomes bound to an external gravitational source. 
The resulting compact object in this case could be sustained by the gravitational field of an external massive body instead of its own self-gravity. We will refer to an object of this kind as a \emph{relaxion halo}. There are significant uncertainties associated with this scenario, which we will return to in future work.
Here, we assume such a halo can exist and investigate the consequences in terrestrial experiments. 

We focus on the relaxion halo hosted by the Sun and by the Earth.
In this case, $M_{\rm ext}$ is either the mass of the Sun or the Earth, and $R_{\rm ext}$ is the corresponding radius. Assuming $M_\star \ll M_{\rm ext}$, the radius of a relaxion halo is
\begin{equation} \label{Rstarext}
R_\star \equiv 
\begin{cases}
\displaystyle{\frac{\Mp^2}{m_\phi^2}\frac{1}{M_{\rm ext}} }& \textrm{for } R_\star > R_{\rm ext}\,,
\\
\displaystyle{\left( \frac{\Mp^2}{m_\phi^2}\frac{R_{\rm ext}^3}{M_{\rm ext}} \right)^{1/4}} & \textrm{for } R_\star \leq R_{\rm ext}\,.
\end{cases}
\end{equation}
The radius of a relaxion halo is determined by the gravitational potential of the external source. 
In the first case, $R_\star > R_{\rm ext}$, we approximate the external source as a point-like mass, which results in an exponential relaxion-halo profile, Eq.~\eqref{profileE}.
In the second case, $R_\star < R_{\rm ext}$, we approximate the external source as a constant-density sphere, where the gravitational potential is given as that of a harmonic oscillator and the profile is Gaussian, Eq.~\eqref{profileG}; though this approximation is rough in principle, in practice it works well when $R_\star \gg R_{\rm ext}$, and to go beyond it is outside the scope of this work. 
See Supplementary Material \ref{app:ASts} for details regarding these two profiles.
Note that in both of these cases, the radius is independent of $M_\star$.
We only consider $M_\star < M_{\rm ext}/2$ for $R_\star > R_{\rm ext}$, and $M_{\star} < (M_{\rm ext}/2) (R_\star / R_{\rm ext})^3$ for $R_{\star} < R_{\rm ext}$, ensuring that the self-gravity is subdominant. 

\begin{figure}[t]
\centering
\includegraphics[scale=0.6]{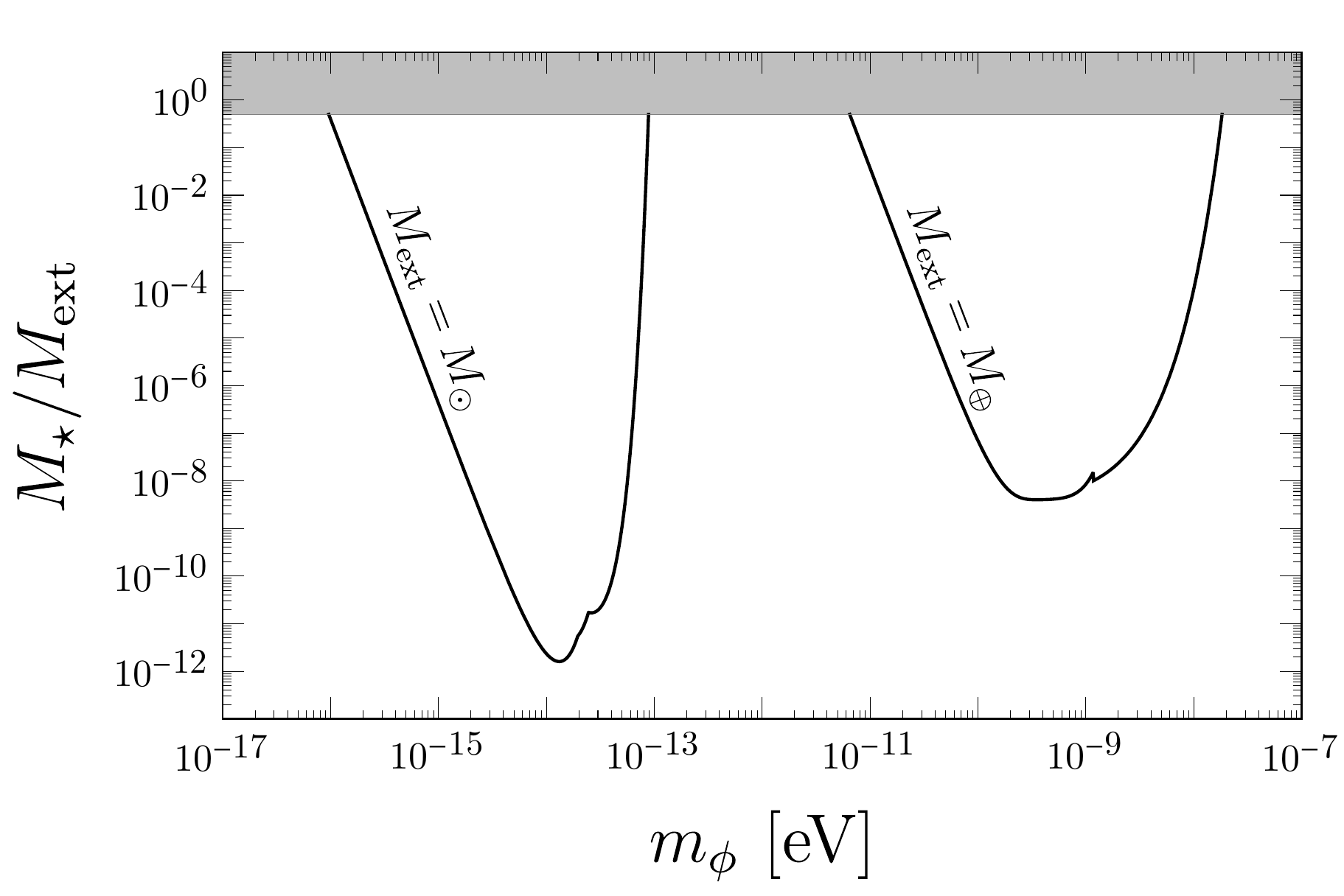}
\caption{%
The upper bound $(M_\star)_{\rm{max}}$ on the relaxion halo mass $M_\star$ as a function of scalar particle mass $m_\phi$; the regions above the black lines are excluded by either (right side, assuming an Earth halo) lunar laser ranging~\cite{Adler:2008rq}, or (left side, assuming a Solar halo) planetary ephemerides~\cite{Pitjev:2013sfa}. We also require $M_\star \leq M_{\mathrm{ext}} / 2$ (boundary of gray shaded region), as explained in the Supplementary Material \ref{app:Constraints}.}
\label{fig:max_Mstar_modified}
\end{figure}

In the presence of an external gravitational source of mass $M_{\rm ext}$, the ground state profile for a relaxion halo is modified compared to the relaxion star. 
To obtain the density and the amplitude of oscillation, we use exponential and Gaussian profiles for $R_\star > R_{\rm ext}$ and $R_\star \leq R_{\rm ext}$ respectively (Eqs.~\eqref{profileE} and \eqref{profileG} in the Supplementary Material). The asymptotic behavior of the halo density is
\begin{equation}
\rho_\star \propto \begin{cases}
\displaystyle{\exp\left(-2r/R_\star\right)}& \textrm{for } R_\star > R_{\rm ext}\,,
\\
\displaystyle{\exp\left(-r^2/R_\star^2\right)} & \textrm{for } R_\star \leq R_{\rm ext}\,.
\end{cases}
\end{equation}
The relevant quantity for experimental searches is the density of relaxion field at the surface of the Earth. We see from Eq.~\eqref{var_const} that the variation of fundamental constants is given by $\delta m_e /\langle m_e\rangle = g_e \sqrt{2 \rho_\star} /( \langle m_e\rangle m_\phi)$ and $\delta \alpha / \alpha = g_\gamma \sqrt{2 \rho_\star} / m_\phi$. 
We discuss various probes to detect these effects in the next section.

One can determine an upper bound on the mass $M_\star$ of a relaxion halo through gravitational observations. In the case of an Earth-based halo, the strongest constraint arises from lunar laser ranging~\cite{Adler:2008rq}, and for a Solar-based halo, from planetary ephemerides~\cite{Pitjev:2013sfa}; both are described in the Supplementary Material \ref{app:Constraints}.\footnote{We consider other possible constraints on an Earth halo in Supplementary Material \ref{app:GPS}, but conclude that~\cite{Adler:2008rq} represents the strongest constraint.}  We show the derived constraint on the mass of a relaxion halo as a function of the scalar particle mass $m_\phi$ in Fig.~\ref{fig:max_Mstar_modified}. Using the result of $(M_\star)_{\rm max}$, we obtain the scalar field value $\phi$, which is directly related to the observables, $\delta m_e/\langle m_e\rangle$ and $\delta \alpha/\alpha$, which we discuss in the next section.

Finally, we comment on the coherence properties of the relaxion halo oscillations.
Because a relaxion halo is supported against collapse by gradient energy, the coherence length of the halo is nothing other than its radius; that is,
\begin{equation}
 R_{\rm coh} = \frac{1}{m_\phi\,v} = \frac{1}{m_\phi}\sqrt{\frac{R_\star\,\Mpl{}^2}{M_{\rm ext}}} = R_\star\,,
\end{equation}
where $v$ is the velocity dispersion in the halo.
The coherence time can be estimated similarly;
for a relaxion Earth halo, we find
\begin{equation}
 \tau_{\rm coh} = \frac{1}{m_\phi\,v^2} = m_\phi\,R_\star^2 \approx
  \begin{cases}
 10^3 \text{ sec}\left(10^{-9} \text{ eV}/m_\phi\right)^3 & \textrm{for } R_\star > R_{\oplus}\,,
\\
 10^3 \text{ sec} & \textrm{for } R_\star \leq R_{\oplus}\,,
\end{cases}
\end{equation}
where we used the radii of Eq.~\eqref{Rstarext} with $M_{\rm ext} = M_\oplus$. For a Solar halo, the coherence time is at least two orders of magnitude larger, as it is enhanced by a large $R_\star \gtrsim 1$ AU in that case.

\begin{figure}[tb]
\centering
\includegraphics[scale=0.46]{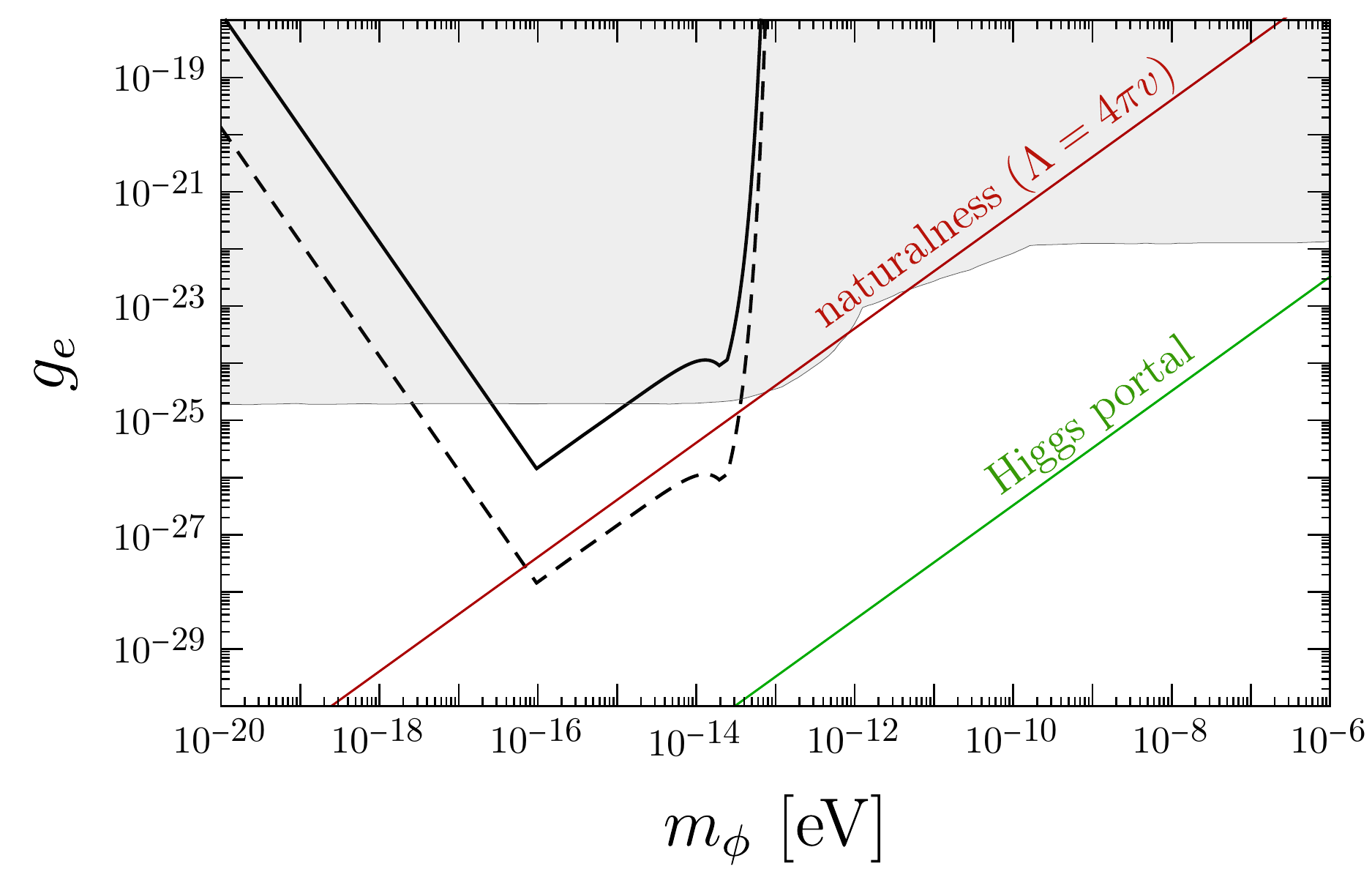}
\,
\includegraphics[scale=0.46]{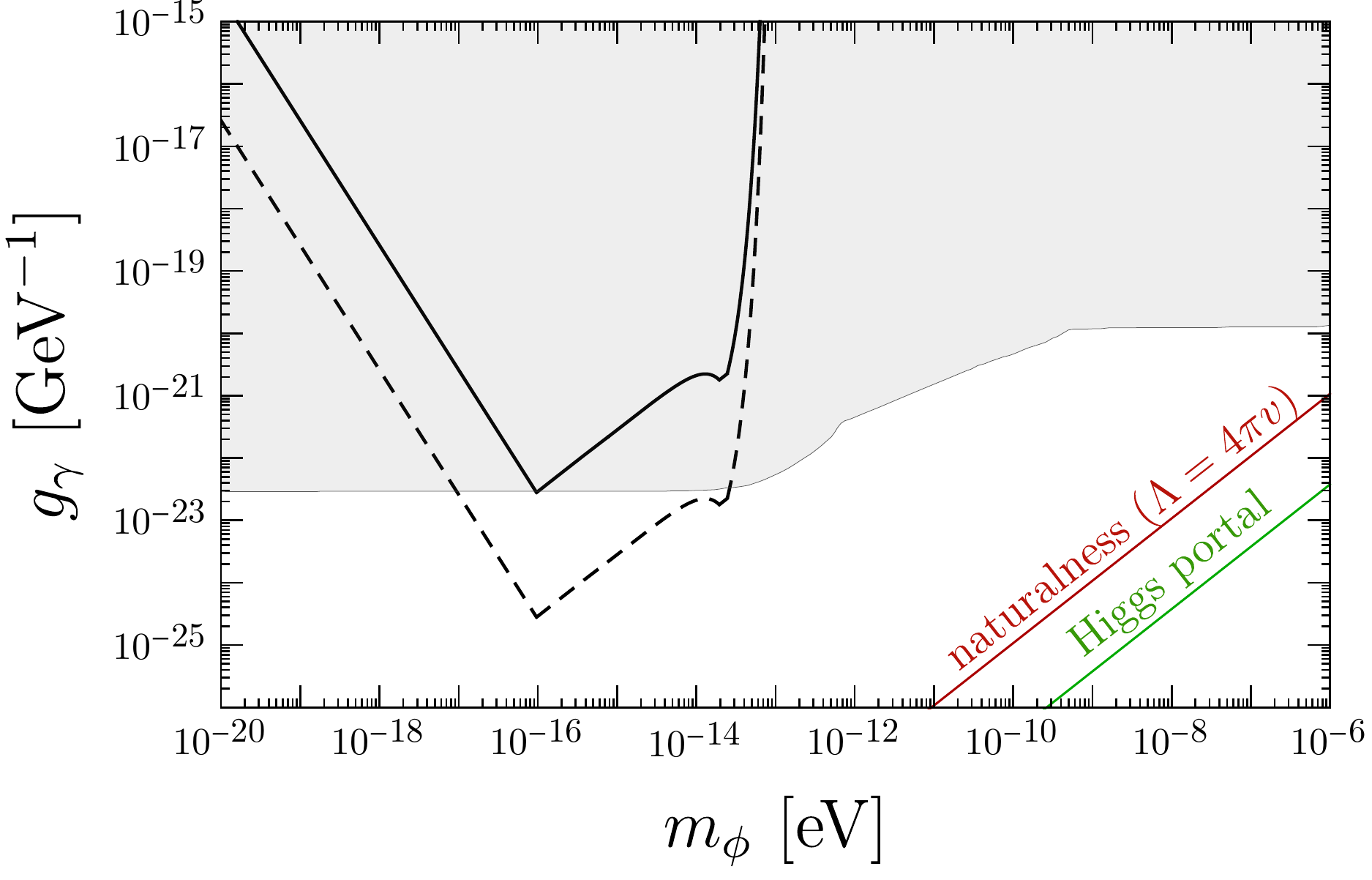}
\caption{Projected constraints on $g_e$ (left) and $g_\gamma$ (right) for a relaxion Solar halo.
Experimental sensitivities in $\delta m_e /\langle m_e\rangle$ and $\delta\alpha/\alpha$ are taken to be  $10^{-16},\,10^{-18}$ (solid and dashed lines, respectively). 
The gray shaded region is excluded by fifth-force experiments. 
The red line is the naturalness limit, where the cutoff is taken to be $\Lambda = 3\TeV$, while the green line is an upper limit on coupling constants which can be obtained from physical relaxion models. 
The halo mass is taken as $M_\star =  \min [(M_{\odot}/2) (R_\star/R_{\odot})^3, (M_\star)_{\rm max}]$, as explained in the Supplementary Material \ref{app:Constraints}.}
\label{fig:ge_Sun}
\end{figure}

\section{Hunting for relaxion halos with table-top experiments} \label{sec:experiments}
As explained above, the possibility of relaxion halos surrounding the Earth or the Sun may lead to an enhanced signal in various table-top experiments. Using the maximally allowed relaxion halo mass as an input, and also using the approximate form of scalar field profile described in Supplementary Material~\ref{app:ASts}, we can compute the oscillation amplitude and compare it to the corresponding experimental sensitivities. 
In order to study the present/near-future sensitivity, we consider the following four cases: 
\begin{itemize}
\item[(i)] Solar-based relaxion halo which is relevant for $m_\phi\sim 10^{-15}\,$eV - bounds on $\delta m_e / \langle m_e\rangle$ and  on $\delta \alpha / \alpha$ are separately considered;
\item[(ii)] Earth-based relaxion halo which is relevant for $m_\phi\sim 10^{-10}\,$eV -  bounds on $\delta m_e / \langle m_e\rangle$
and on $\delta \alpha / \alpha$ are separately considered.
\end{itemize}

\begin{figure}[t]
\includegraphics[scale=0.46]{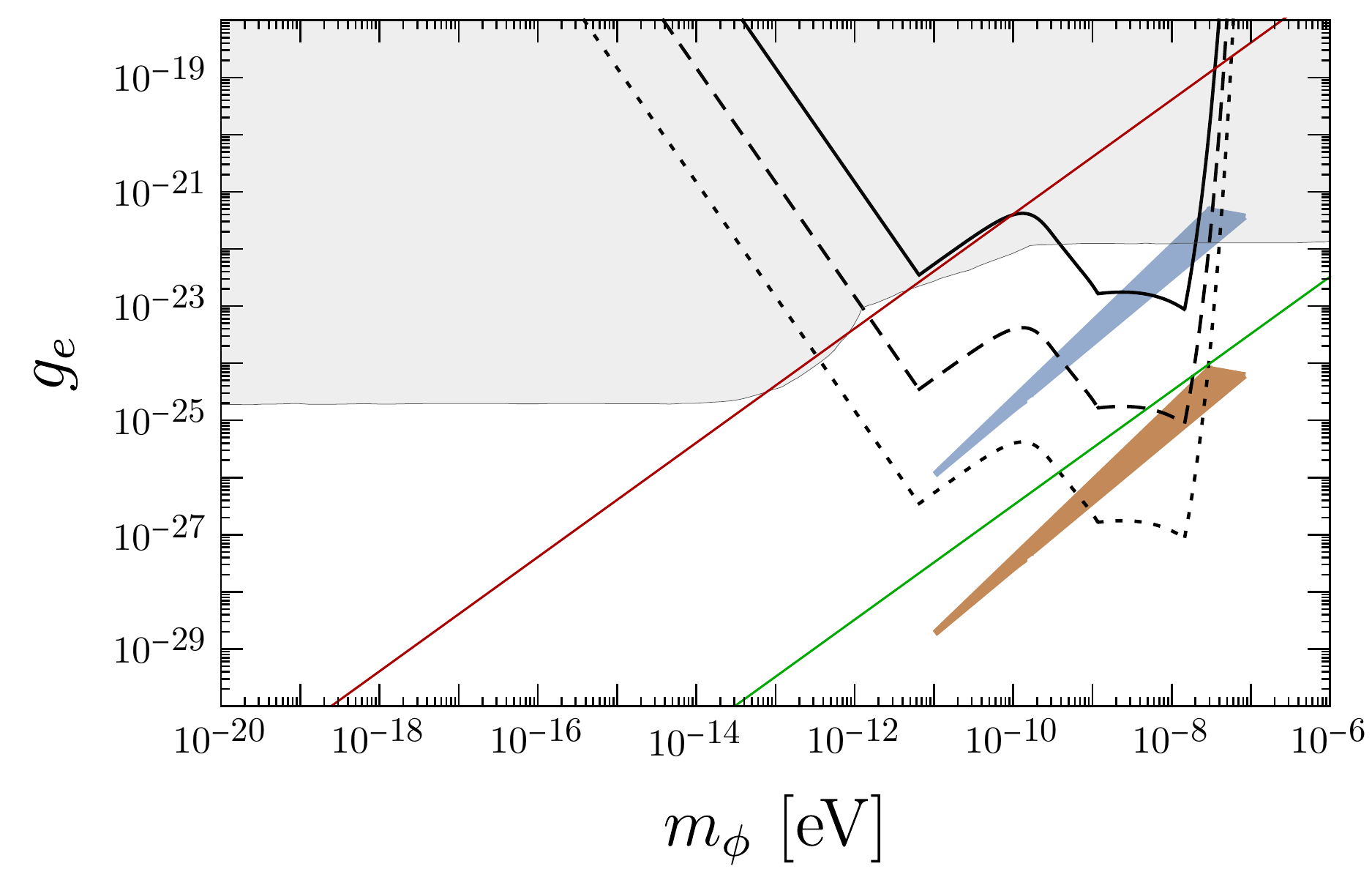}
\,
\includegraphics[scale=0.46]{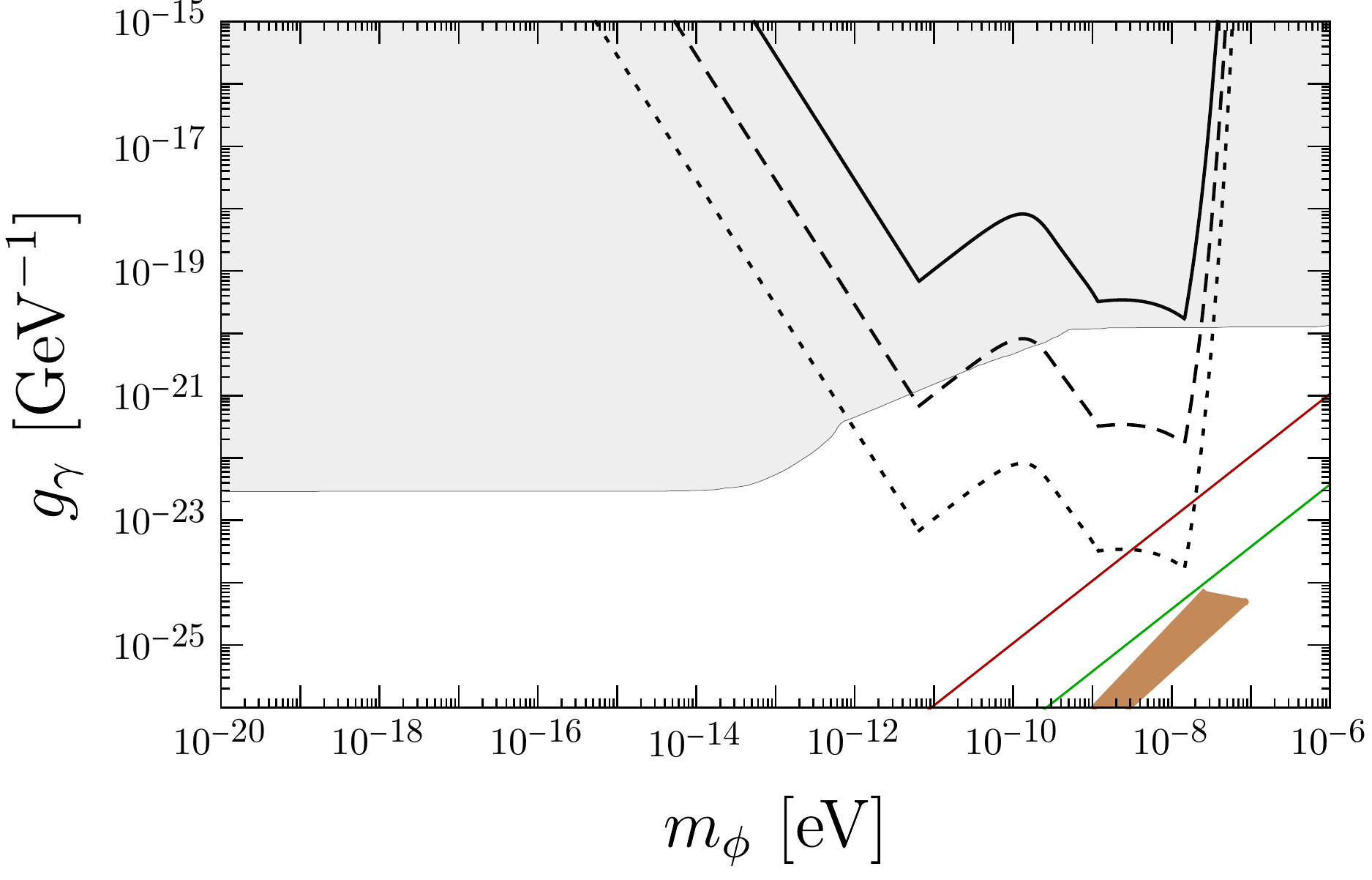}
\caption{Projected constraints on $g_e$ (left) and $g_\gamma$ (right) for a relaxion Earth halo.
Experimental sensitivities in $\delta m_e /\langle m_e\rangle$ and $\delta\alpha/\alpha$ are taken to be $10^{-14},\,10^{-16},\,10^{-18}$ (solid, dashed, and dotted lines, respectively). The gray shaded regions, as well as the red and green lines, are the same as in Fig. \ref{fig:ge_Sun}.
The halo mass is taken as $M_\star =\min [(M_{\oplus}/2) (R_\star/R_{\oplus})^3, (M_\star)_{\rm max}]$, as explained in the Supplementary Material \ref{app:Constraints}. The shaded regions represent the allowed regions for coherent relaxion DM \cite{Banerjee:2018xmn}, as explained in the text.}
\label{fig:ge_Earth}
\end{figure}

For case (i), we show in  Fig.~\ref{fig:ge_Sun} sensitivity curves for $(\delta m_e / \langle m_e\rangle,\,\delta \alpha/\alpha) = 10^{-16}$\,(solid lines) and $10^{-18}\,$ (dashed lines). In addition, the bounds from fifth-force and equivalence-principle tests correspond to the shaded region~\cite{Touboul:2017grn,Berge:2017ovy,Schlamminger:2007ht,Smith:1999cr}, the red line corresponds to the naturalness limit with a cutoff at $ \Lambda = 3\TeV$ (the minimal allowed cutoff consistent with solving the hierarchy problem), and the green line corresponds to the naive upper limit on coupling constants which derived from scalar-Higgs portal models \cite{Piazza:2010ye}.
Note that the bounds from equivalence-principle tests are obtained by neglecting the other possible couplings of scalar field to SM particles.

In Fig.~\ref{fig:ge_Earth}, we show the analogous sensitivities in case (ii), with $(\delta m_e / \langle m_e\rangle,\,\delta \alpha/\alpha) = 10^{-14}$\,(solid), $10^{-16}\,$ (dashed) and $10^{-18}\,$ (dotted).
In the case of a Solar halo, future projections for $g_e$ reach the parameter space where the scalar mass is technically natural, while in the case of an Earth halo, future projections reach not only to the naturalness limit for $g_e$ and $g_\gamma$, but also to the region of physically motivated generic relaxion models~\cite{Flacke:2016szy,Frugiuele:2018coc}. The shaded regions represent the allowed parameter space for coherent relaxion dark matter \cite{Banerjee:2018xmn}, taking $g_e = y_e\,\sin\theta$ and $g_\g = (\a / 4\pi v)\,\sin\theta$; in the left panel, the brown region is for $y_e = y_e^{\rm SM}$ (the Standard Model prediction), whereas the blue region is for $y_e = 600\times y_e^{\rm SM}$, the maximum allowed value given current LHC constraints \cite{Dery:2017axi}.

\section{Outlook}

In this work, we consider the effect of (pseudo)scalar field dark matter, e.g. relaxion dark matter, in atomic physics experiments. We propose that such dark matter can form gravitationally bound objects denoted as boson stars (or relaxion stars), and suggest that these stars can be formed around the Earth or the Sun leading to relaxion halos with density well above that of the local DM. Due to the mixing with the Higgs, the oscillating DM background implies that all the fundamental couplings of nature are varying with time. This implies that one could search for signals of such objects in table-top experiments, which may be probed in the near future with projected sensitivity stronger than that of fifth-force and equivalence-principle tests. In this scenario, even present experimental sensitivity \cite{Aharony:2019iad,Antypas:2019qji} may be sufficient to probe the parameter space of coherent relaxion DM \cite{Banerjee:2018xmn}.

We note that as our signal is related to rapid-oscillation signals, other existing probes of scalar DM, which are DC-oriented and/or using less precise clocks~\cite{Derevianko:2013oaa,Roberts:2017hla,Wolf:2018xlz,Derevianko:2016vpm} would be less sensitive to the above form of DM. However, in the case of a relaxion halo or star which coherently oscillates over sufficiently large distances one may improve the sensitivity to its presence by comparing the phase of the oscillation between two distant experiments (or network of sensors) that are synched to the same external clock, or similarly if a single experiment is to repeat its measurements multiple times while being synchronised to an external clock. Furthermore, there are several proposals for sending high performance clock-systems to space~\cite{Schiller:2012qn,Kolkowitz:2016wyg}, which would allow to map the relaxion halo density as a function of distance from the Earth's surface.

Another interesting implication is the possible presence of mini-relaxion halos, whose radius is smaller than that of the Earth so that such halos do not contribute to the signals described above.
Such objects arise when the relaxion particle mass is around nano-eV or above.
Although they can have densities close to that of the Earth, it is in general difficult to probe them because they are located beneath the surface of the Earth (see however~\cite{PhysRevD.86.107502}, which proposes to test clock universality in deep underground/underwater experiments).

We finally reiterate that our discussion and conclusion throughout the paper holds for any form of light scalar dark matter, thus covering a large parameter space of well-motivated dark matter models. 

\subsection*{Acknowledgements}
We are grateful for useful discussions and comments on the manuscript from Kfir Blum, Itay Halevy, Eric Kuflik, Mordehai Milgrom, Roee Ozeri, Gil Paz, Stephan Schiller, L.C.R. Wijewardhana, and Hong Zhang. 
The work of DB is supported by the European Research Council (ERC) under the European Unions Horizon 2020 research and innovation programme (grant agreement No 695405), the Simons and Heising-Simons Foundations and the DFG Reinhart Koselleck project. 
The work of JE is supported by the Zuckerman STEM Leadership Program.
The work of GP is supported by grants from the BSF, ERC, ISF, Minerva, and the Segre Research Award.

\providecommand{\href}[2]{#2}\begingroup\raggedright\endgroup

\newpage

\setcounter{page}{1}

\section*{{\Large Supplementary Material} \\ \vspace{0.3cm} Abhishek Banerjee, Dmitry Budker, Joshua Eby, Hyungjin Kim, and Gilad Perez}
{\begin{center} 
In this Supplementary Material, we give further details necessary to derive the constraints of the main text. In particular, we describe in detail the density profiles used for relaxion stars and halos, and the approximations that are appropriate in each case. We also analyze the constraints from gravitational measurements used to determine the maximum relaxion halo mass, both for the Earth-based and Solar-based cases.
\end{center}}

\beginsupplement
\section{Stable configurations of scalar particles} \label{app:ASts}
In this section, we review some important properties of self-gravitating relaxion stars, and define the approximate density profiles used in the main text to describe both relaxion stars and halos.

We neglect the effect of self-interactions in this discussion, but comment on this topic at the end of this section.
In the nonrelativistic limit, it is convenient to introduce a relaxion wavefunction $\psi$,
\bea
\phi = \frac{1}{\sqrt{2m_\phi}} ( \psi \,e^{-im_\phi t} + \psi^* e^{im_\phi t} ).
\eea
The equation of motion describing the relaxion wavefunction is
\begin{align}
 i\,\pd_t\,\psi &= \left[-\frac{\nabla^2}{2m_\phi} + V_g(|\psi|) + V_{\rm ext}(R_{\rm ext},M_{\rm ext})\right]\psi, 
\end{align}
where
\begin{equation}
 V_g = - G\,m_\phi^2\,\int d^3r' \frac{|\psi(r')|^2}{|\vec{r} - \vec{r}\,'|}
\end{equation}
is the Newtonian gravitational potential of the relaxion star, and $V_{\rm ext}$ is that of an external source of gravitation. Note that the normalization of the wavefunction is determined by the requirement $\int d^3r |\psi(r)|^2 = M_\star/m_\phi$.

In the case of a self-gravitating relaxion star (Section \ref{sec:stars}), we can set $V_{\rm ext}=0$, and analyze stable configurations by minimizing the energy functional
\begin{align}
 E = \int d^3r \left[\frac{|\nabla\psi|^2}{2m_\phi} + \frac{1}{2}V_g\,|\psi|^2 \right]
 = \frac{A\,M_\star}{m_\phi^2\,R_\star^2} - \frac{B\,M_\star^2}{\Mp^2\,R_\star}.
\end{align}
We use a linear$\times$exponential ansatz for the wavefunction (providing a good fit \cite{Eby:2018dat}) of the form
\begin{equation}
 \psi(r) = \sqrt{\frac{M_\star}{7\pi\,m_\phi\,R_\star^3}} \left(1 + \frac{r}{R_\star}\right) \exp\left(-r/R_\star\right)\,,
\label{app_profile}
\end{equation}
which determines the dimensionless coefficients
\begin{align}
A &= \frac{3}{14} \approx 0.2\,, 
&B& = \frac{5373}{25088} \approx 0.2\,. 
\end{align}
In this case we recover the standard mass-radius relation of a stable boson star, Eq.~\eqref{Rstar}, which is well known (see {\it e.g.} \cite{Chavanis:2011zi}).

For the relaxion halo (Section \ref{sec:halo}), we neglect $V_g$ but must consider two important limits: when $R_\star\gg R_{\rm ext}$, we can approximate $V_{\rm ext}$ by the potential of a point particle of mass $M_{\rm ext}$; and in the other limit $R_\star\ll R_{\rm ext}$, we approximate the Earth or the Sun as a constant-density sphere. Thus we use the potential
\begin{equation} \label{Vext}
 V_{\rm ext} =
 \begin{cases}
-\displaystyle{\frac{G\,m_\phi\,M_{\rm ext}}{r}} & \textrm{for } R_\star > R_{\rm ext}\,,
\\
- \displaystyle{\frac{3\,G\,m_\phi\,M_{\rm ext}}{2\,R_{\rm ext}}\left[1 - \frac{1}{3}\left(\frac{r}{R_{\rm ext}}\right)^2\right]} & \textrm{for } R_\star \leq R_{\rm ext}\,,
\end{cases}
\end{equation}
in these calculations.
In the first case, the system is essentially a gravitational atom which is well-described by an exponential profile
\begin{equation} \label{profileE}
 \psi(r) = \sqrt{\frac{M_\star}{\pi\,m_\phi\,R_\star^3}} \exp\left(-r/R_\star\right) 
 		\quad \textrm{ for } M_\star \ll M_{\rm ext} \textrm{ and } R_\star > R_{\rm ext}\,,
\end{equation}
where $R_\star$ is given by the top line of Eq.~\eqref{Rstarext}. On the other hand, at larger values of $m_\phi$ we can have $R_\star < R_{\rm ext}$, in which case the potential in the inner region is that of a three-dimensional isotropic harmonic oscillator; the solution is a Gaussian function
\begin{equation} \label{profileG}
 \psi(r) = \sqrt{\frac{M_\star}{\pi^{3/2}\,m_\phi\,R_\star^3}} \exp\left(-\frac{1}{2}\frac{r^2}{R_\star^2}\right) 
 		\quad \textrm{ for } M_\star \ll M_{\rm ext} \textrm{ and } R_\star < R_{\rm ext}\,,
\end{equation}
where $R_\star$ is given by the bottom line of Eq.~\eqref{Rstarext}. To minimize errors in the intermediate range $R_\star = \Ocal(R_{\rm ext})$, we use the Gaussian function in the inner region (for $r < R_{\rm ext}$) and match at the boundary to an exponential profile for $r\geq R_{\rm ext}$. In this procedure, we somewhat underestimate the mass contained in the tail of the relaxion halo density function when $R_\star < R_{\rm ext}$, leading to an error in the constraint on $M_\star$ in Fig. \ref{fig:max_Mstar_modified} which is no larger than a factor of $2$ at any $m_\phi$ we consider. These profiles determine the density of the relaxion halo, which we used in Sections \ref{sec:halo} and \ref{sec:experiments}.

We now comment on the effect of self-interactions. Suppose the self-interaction potential for the scalar field $\phi$ is \cite{Banerjee:2018xmn2}
\begin{equation} \label{Vphi}
 V(\phi) = \L_{\rm br}^4 \left[1 - \cos\left(\frac{\phi}{f}\right)\right].
\end{equation}
Here, $\L_{\rm br}$ is related to the scale where the potential of $\phi$ is generated, and the scalar mass is $m_\phi = \L_{\rm br}^2/f$ for decay constant $f$. 
The leading-order self-interaction in the expansion of Eq.~\eqref{Vphi} is proportional to $-(m_\phi^2/f^2)\phi^4$. In the standard case of a boson star ($M_{\rm ext} \to 0$), this interaction  triggers gravitational instability for $M_\star$ greater than a critical value \cite{Chavanis:2011zi,Eby:2014fya}
\begin{equation} \label{Mcritic}
 M_{c,0} \approx 10\,\frac{\Mp f}{m_\phi}\,,
\end{equation}
meaning that for $M_\star > M_{c,0}$ a relaxion star is unstable and will collapse. Furthermore, the mass-radius relation is modified when $M_\star$ is close to $M_{c,0}$ as well.
In the case of an external gravitational potential, the critical mass becomes smaller, implying instability whenever $M_\star$ is larger than
\begin{equation}
 M_c \approx \frac{M_{c,0}^2}{M_{\rm ext}}.
\end{equation}
These effects are not relevant over most of the parameter space we consider here, as the requirement of stability amounts to a rather modest constraint on the decay constant, $f \gtrsim \Ocal\big(10^7\big)$ GeV at the strongest. The purple shaded regions of Fig.~\ref{fig:transient} represent this unstable region.

\section{Constraints on relaxion halos from local gravity measurements} \label{app:Constraints}

In this section, we derive the constraint on the relaxion halo mass $M_\star$ for both the Earth-based and Solar-based cases, as a function of the relaxion particle mass $m_\phi$.

\begin{figure}
\centering
\includegraphics[scale=0.6]{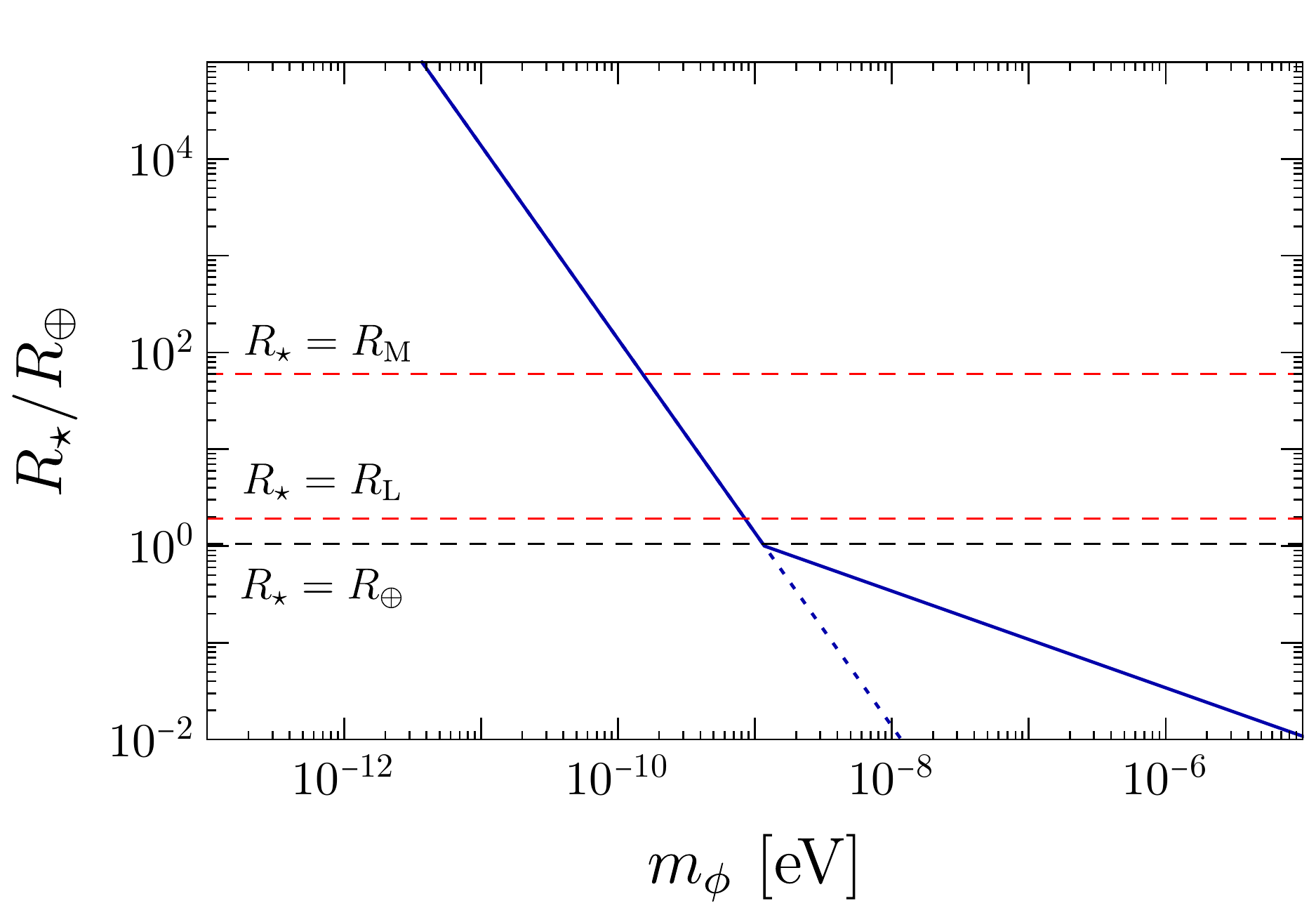}
\caption{The radius $R_\star$ of a relaxion halo which is supported by the Earth's gravity, as a function of the scalar particle mass $m_\phi$; the solid blue line is determined from Eq.~\eqref{Rstarext} assuming the Earth is a uniform-density sphere, whereas the dashed blue is found when treating the Earth as a pointlike source. The red dashed lines denote the radii of the Moon's orbit $R_M \approx 60R_\oplus$ and the LAGEOS satellite $R_L \approx 2R_\oplus$; see text for details.
}
\label{fig:r_vs_m}
\end{figure}

The maximum allowed relaxion halo mass $M_\star$ depends importantly on the radius of the halo, given in Eq.~\eqref{Rstarext}. In Fig.~\ref{fig:r_vs_m}, we illustrate the radius of a relaxion halo as a function of the relaxion mass $m_\phi$ (solid blue).
We also present the radius of the Earth (black dashed line), and radii of the orbits of the Moon and the LAGEOS satellite (red dashed lines), which are relevant for discussion of astrophysical constraint on $M_\star$. The transition from the point-like approximation (of the external gravitational potential) to the constant-density approximation occurs at $m_\phi \approx 1.2\times10^{-9}$ eV.

The total mass of a relaxion halo is constrained by various astrophysical observations, as it may change the motion of astrophysical bodies around the solar system, or objects in orbit around the Earth.
If the relaxion halo is hosted by the Earth, the strongest constraints arises from lunar laser ranging and the LAGEOS satellite~\cite{Adler:2008rq2}, which constrains the total mass lying between the orbits of the Moon and the satellite to be $M_{\rm enc}< M_{\rm enc}^{\rm max} = 4\times 10^{-9} M_\oplus$.
The mass enclosed between these two orbits from a relaxion halo is
\bea
M_{\rm enc} = 4\pi\,M_\star \int_{R_{\rm L}}^{R_{\rm M}} dr\,  r^2 \frac{\rho_\star(r)}{M_\star} ,
\eea
where the orbital radius of the Moon is $R_{\rm M} \simeq 60 R_\oplus$, while that of the LAGEOS satellite is $R_{\rm L} \simeq 2 R_\oplus$. The quantity in the integral does not depend on $M_\star$, but only on the shape of the wavefunction in the region of integration.
This leads to the upper bounds on the mass of relaxion halo hosted by the Earth as
\bea
(M_\star)_{\rm max} = \frac{M_{\rm enc}^{\rm max}}{4\pi} 
\left[  \int_{R_{\rm L}/R_\star}^{R_{\rm M}/R_\star} dx\,x^2\, \frac{R_\star^3\,\rho_\star(x)}{M_\star} \right]^{-1},
\eea
where $x=r/R_\star$.

Similarly, the mass of a Sun-based halo is constrained by planetary ephemerides because, for $m_\phi<10^{-14}\eV$, we have $R_\star \gtrsim 1 \, {\rm AU}$, and thus this extended mass distribution can change the orbital motion of planets.
In Ref.~\cite{Pitjev:2013sfa2}, the dark matter density at the orbital radii of planets in our solar system (Mercury, Venus, Earth, Mars, Jupiter, and Saturn) are constrained to $\rho \lesssim 10^{-18} \,\textrm{--}\, 10^{-20} \,{\rm g/cm}^{3}$.
We use Eq.~\eqref{profileE} to compute the density of the relaxion Solar halo at the orbital radius of each planet, translate the result of~\cite{Pitjev:2013sfa2} as upper limits on $M_\star$, and take the strongest constraint as the limiting value of $M_\star$. 

\begin{figure}
\centering
\includegraphics[scale=0.49]{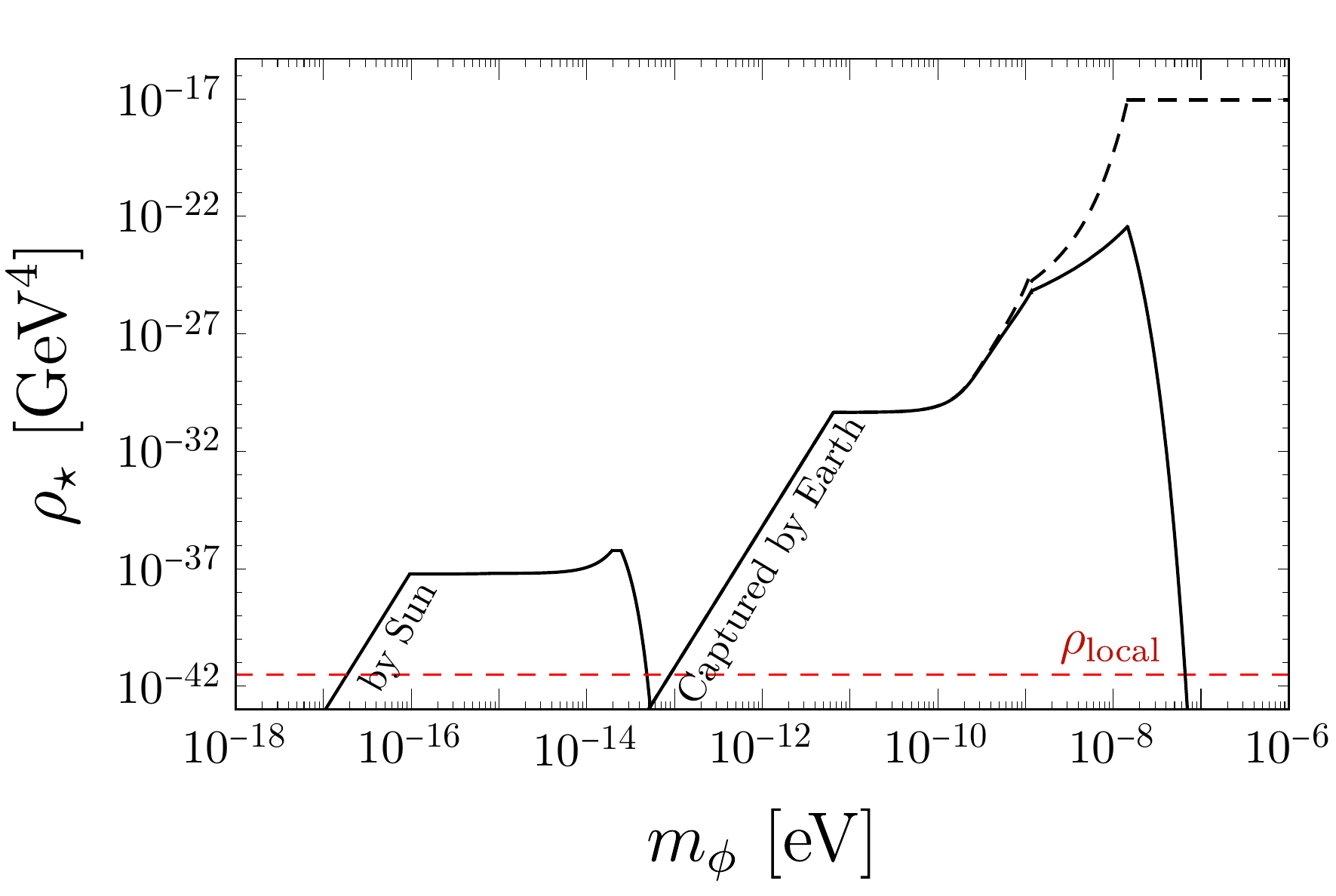}
\,
\includegraphics[scale=0.465]{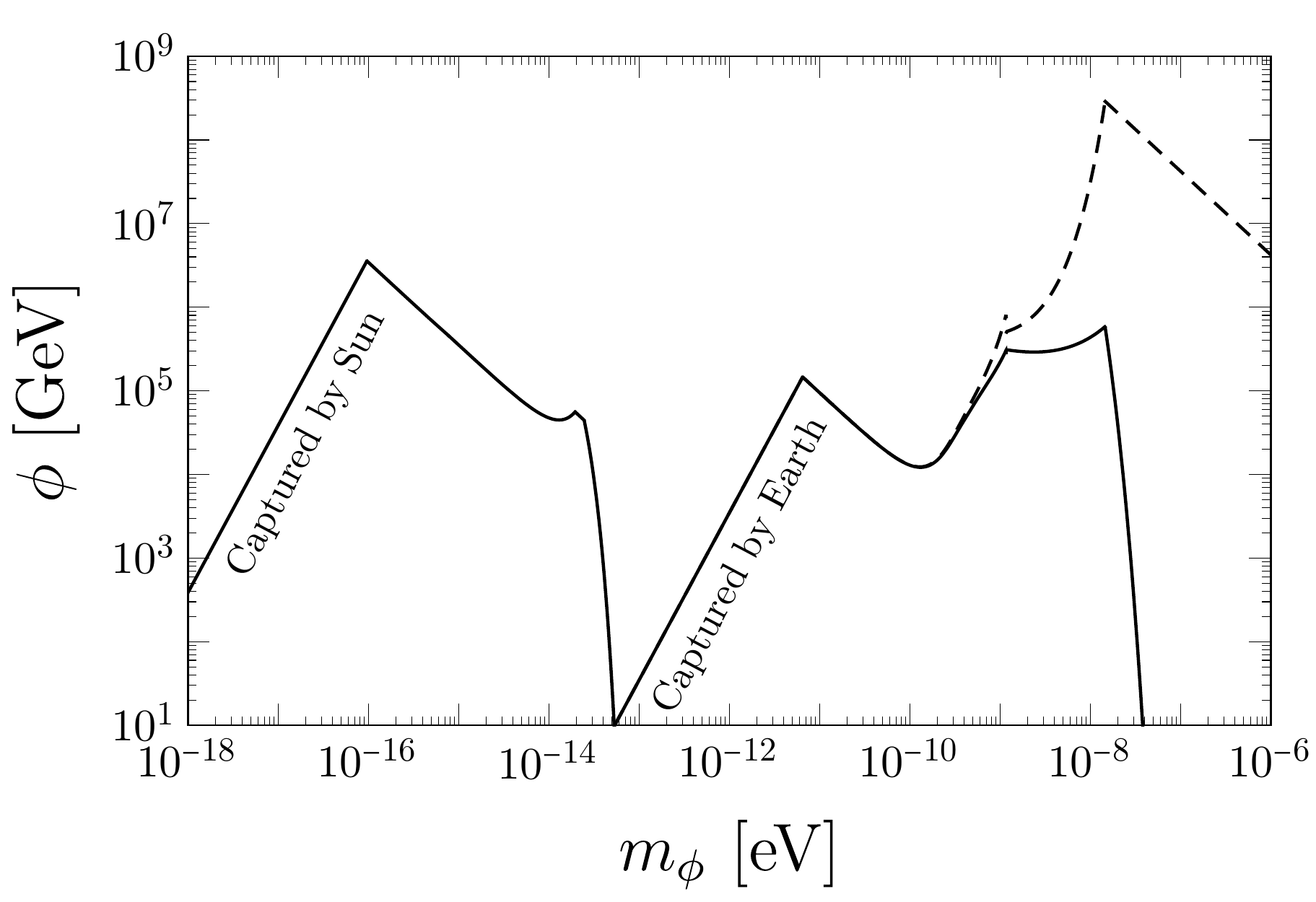}
\caption{Density of relaxion halo $\rho_\star$ (left) and corresponding scalar field value $\phi$ (right). 
The solid line represents the density and the field value of relaxion halo at the surface of the Earth, while the dashed line is the value at the center of the relaxion halo.
For this plot, we take $M_\star = \min [(M_{\rm ext}/2) (R_\star/R_{\rm ext})^3, (M_\star)_{\rm max}]$, where $(M_\star)_{\rm max}$ is shown in Fig. \ref{fig:max_Mstar_modified}.}
\label{fig:rho_and_phi}
\end{figure}

In Fig.~\ref{fig:max_Mstar_modified}, we show the upper limit on $M_\star$ as a function of $m_\phi$.
The black solid line on the right side is the constraint on the mass of a relaxion Earth halo, obtained from lunar laser ranging~\cite{Adler:2008rq2}.
The constraint is most severe when $R_{\rm L} < R_\star < R_{\rm M}$, which can be translated into $10^{-10}\eV\lesssim m_\phi \lesssim 10^{-9}\eV$. 
This is because, for a fixed $M_\star$, it is this range of $R_\star$ that most of the mass is contained in $r\in [R_{\rm L}, R_{\rm M}]$. 
As $m_\phi$ increases, the relaxion star becomes increasingly concentrated on $R_\star< R_{\rm L}$;
as $m_\phi$ decreases, the size of the relaxion halo is even larger than the orbital radius of Moon, and thus only small fraction of its total mass is enclosed in $R\in [R_{\rm L}, R_{\rm M}]$, so that $M_\star$ is less constrained.
On the other hand, the black solid line on the left side is obtained from planetary ephemerides in the case of a Solar halo~\cite{Pitjev:2013sfa2}.
Similarly, the constraint is the most severe when $R_\star$ is order of a few AU. As described at the start of Section \ref{sec:halo}, we further limit the mass of any relaxion halo to $M_\star < M_{\rm ext}/2$, bounding the gray shaded region in Fig.~\ref{fig:max_Mstar_modified}.

This upper limit on $M_\star$ can be directly translated into the density of a relaxion halo and the scalar field value at the surface of the Earth, which are directly relevant for the observables.
Using the result of $(M_\star)_{\rm max}$ above, we present the density (left panel) and the field amplitude (right panel) of the relaxion halo computed at the surface of the Earth as the solid lines in Fig.~\ref{fig:rho_and_phi}.
For the mass of the relaxion halo, we choose
\bea
M_\star = \min [ (M_\oplus /2)(R_\star/R_\oplus)^3, (M_\star)_{\rm max}],
\eea
ensuring that the structure of relaxion halo is maintained by the external gravitational field and is consistent with gravitational constraints. 
In the figure, the black dashed line indicates the density and field value of the Earth halo at the center of the Earth. 
The local DM density, $\rho_{\rm local} = 0.4\GeV/{\rm cm}^3$ is given by the red dashed line.
The scalar field value $\phi$ obtained is directly related to the observables, $\delta m_e/\langle m_e\rangle$ and $\delta \alpha/\alpha$.

\section{Possible constraint on relaxion halo mass from GPS satellites} \label{app:GPS}

Additional mass $M_\star$ coming from a relaxion halo would change the gravitational field around the Earth.
In the absence of an independent measurement of the total mass of the Earth, the only constraint on the mass of relaxion star is, perhaps, from measurements of a deviation from the gravitational inverse-square law.
For instance, in~\cite{Speake:1990cq}, the gravitational acceleration, ``little $g$" is measured at different tower heights up to $300$ m, and with accuracy of $\Delta g /g \sim 10^{-8}$ to test the inverse-square law of Newtonian gravity.
The contribution to little $g$ due to a relaxion halo is given as
\begin{equation}
g_\star = \frac{G M_\star }{r^2}\left(\frac{r}{R_\star} \right)^3,
\end{equation}
assuming that the mass density of relaxion star is constant inside its radius.
At the surface of the Earth, the relaxion star changes the gravitational acceleration $g$ by $\Delta g /g = (M_\star / M_\oplus) (R_\oplus / R_\star)^3$, but this shift, again, may not be measurable since one can shift the total mass of Earth in such a way as to cancel the effect. 

However, certainly an anomalous behaviour of $g_\star$ as a function of radius $r$ can be measured if it is large enough. 
The anomalous change of little $g$ due to a relaxion halo over $\Delta r \simeq 300$ m is
\begin{equation}
\Delta g_\star / g \simeq \frac{M_\star}{M_\oplus}\left(\frac{R_\oplus}{R_\star} \right)^2 \frac{\Delta r}{R_\star}
\simeq 10^{-8} \times \left(\frac{M_\star / M_\oplus}{10^{-3}}\right) \left( \frac{R_{\oplus}}{R_\star}\right)^3\,,
\end{equation}
assuming the relaxion halo is larger than the Earth.
Thus, the inverse-square law test constrains the relaxion star mass as $M_\star < 10^{-3} M_\oplus$ when $R_\star = R_\oplus$~\cite{Speake:1990cq}. 
This is weaker than the constraint obtained from~\cite{Adler:2008rq2}, which we have discussed in Section \ref{app:Constraints}.

A more plausible way to put a constraint on the mass of relaxion star is through gravitational redshifts of clock transition frequencies.
Far away from Schwarzschild radius, the gravitational redshift is
\begin{equation}
z \simeq \frac{GM}{r},
\end{equation}
where $M = M_\oplus + M_\star ( r/R_\star)^3$ for $r<R_\star\,$. 
Consider two clocks at radii $r_1$ and $r_2$. 
If $r_1 = r_2$, then the effect of gravitational redshift is the same for both clocks so that we would not be able to see the effect of a relaxion star.
So the net effect arises only when $\Delta r = |r_1 -r_2| \neq 0$, and it is proportional to
\begin{equation}
\Delta z = GM_\oplus \left( \frac{1}{r_1} - \frac{1}{r_2} \right) + \frac{GM_\star}{R_\star^3} (r_1 + r_2)(r_1 - r_2)\,.
\end{equation}
The first term is the usual gravitational redshift due to Earth mass, while the second term is due to the relaxion halo.
The anomalous shift in the transition frequency due to the halo is thus 
\begin{align}
(\Delta z)_\star &= \frac{GM_\star}{R_\star^3} (r_1 + r_2)(r_1 - r_2) \nn \\
	& \simeq 10^{-16} \left( \frac{M_\star}{10^{-5} M_\oplus} \right) \left( \frac{10 R_\oplus}{R_\star} \right)^3
\left(\frac{\Delta r}{3R_\oplus} \right).
\end{align}
Here we assume one clock on Earth, and the other one on GPS satellite such that $\Delta  r \simeq 3 R_\oplus\,$.
For a clock sensitivity of $10^{-16}$, we should be able to see anomalous shift in clock transition frequency for $M_\star / M_\oplus = 10^{-5}$ and for $R_\star = 10 R_\oplus\,$. This, too, is weaker than the constraint obtained from~\cite{Adler:2008rq2}.

\end{document}